\documentclass[default]{sn-jnl}
\usepackage{amsmath}   
\usepackage{ amssymb }
\usepackage[utf8]{inputenc}
\usepackage[T1]{fontenc}
\usepackage{footnote}
\usepackage[english]{babel} 
\usepackage{hyperref}         
\usepackage{graphicx}         
\usepackage{epstopdf}	      
\usepackage{subfig}
\usepackage{bm} 				
\usepackage{booktabs}
\usepackage{color}
\usepackage{float}
\usepackage{upgreek}
\usepackage[nolist]{acronym}

\newcommand{\+}[1]{\text{#1}}

\newlength\fwidth
\newlength\swidth

\newcommand{\qscn}{\,{;}}
\newcommand{\zJ}{\bm J}

\newcommand{\zn}{\bm n}
\newcommand{\zr}{\bm r}
\newcommand{\qpt}{\,{.}}
\newcommand{\qcm}{\,{,}}
\newcommand{\idx}[1]{_\text{#1}}
\newcommand{\udV}{\,\text{d}\Omega}

\newcommand{\pd}[2]{\frac{\partial #1}{\partial#2}}
\newcommand{\dd}[2]{\frac{\text d #1}{\text d#2}}

\usepackage{aligned-overset}
\newcommand{\divergence}[1]{\,\text{div}\left(#1\right)}

\newcommand{\intV}[1]{ \int_\Omega #1 \udV}
\newcommand{\intT}[1]{\int_0^T #1 \+dt}

\newcommand{\grad}[1]{\,\text{grad}\left(#1\right)}

\allowdisplaybreaks

\jyear{2021}%

\theoremstyle{thmstyleone}%
\theoremstyle{thmstyletwo}%

\theoremstyle{thmstylethree}%

\begin{document}

\title[Adjoint Method for Nonlinear EQS Problems]{Adjoint Variable Method for Transient Nonlinear Electroquasistatic Problems}


%
%
%
%

\author*[1,2]{\fnm{M. Greta} \sur{Ruppert}}\email{ruppert@temf.tu-darmstadt.de}

\author[1,2]{\fnm{Yvonne} \sur{Späck-Leigsnering}}\email{spaeck@temf.tu-darmstadt.de}

\author[1]{\fnm{Julian} \sur{Buschbaum}}\email{julian\textunderscore johannes.buschbaum@tu-darmstadt.de}

\author[1,2]{\fnm{Herbert} \sur{De Gersem}}\email{degersem@temf.tu-darmstadt.de}

\affil[1]{\orgdiv{Institute for Aceelerator Science and Electromagnetic Fields (TEMF)}, \orgname{Technische Universität Darmstadt}, \orgaddress{\street{Schloßgartenstraße~8}, \city{64289 Darmstadt}, \country{Germany}}}
\affil[2]{\orgdiv{Graduate School of Computational Engineering}, \orgname{Technische Universität Darmstadt}, \orgaddress{\street{Dolivostraße~15}, \city{64293 Darmstadt}, \country{Germany}}}

%


\abstract{
	Many optimization problems in electrical engineering consider a large number of design parameters. A sensitivity analysis identifies the design parameters with the strongest influence on the problem of interest. This paper introduces the adjoint variable method as an efficient approach to study sensitivities of nonlinear electroquasistatic problems in time domain. In contrast to the more common \acl{dsm}, the \acl{am} has a computational cost nearly independent of the number of parameters. The method is applied to study the sensitivity of the field grading material parameters on the performance of a 320\,kV cable joint specimen, which is modeled as a \acl{fe} nonlinear transient \acl{eqs} problem. Special attention is paid to the treatment of quantities of interest, which are evaluated at specific points in time or space. It is shown that shown that the method is a valuable tool to study this strongly nonlinear and highly transient technical example.}



\begin{acronym}
	\acro{rhs}[RHS]{right-hand side}
	\acrodefplural{rhs}[RHSs]{right-hand sides}
	\acro{2d}[2D]{two-dimensional}
	\acro{3d}[3D]{three-dimensional}
	\acro{ac}[AC]{alternating current}
	\acro{am}[AVM]{adjoint variable method}
	\acro{dc}[DC]{direct current}
	\acro{dof}[DoF]{Degrees of Freedom}
	\acro{dsm}[DSM]{direct sensitivity method}
	\acro{epdm}[EPDM]{ethylene propylene diene monomer}
	\acro{eqs}[EQS]{electroquasistatic}
	\acro{eqst}[EQST]{electroquasistatic-thermal}
	\acro{es}[ES]{electrostatic}
	\acro{fe}[FE]{Finite Element}
	\acro{fem}[FEM]{Finite Element Method}
	\acrodefplural{fgm}[FGMs]{field grading materials}
	\acro{fgm}[FGM]{field grading material}
	\acro{hv}[HV]{high voltage}
	\acro{hvac}[HVAC]{high voltage alternating current}
	\acro{hvdc}[HVDC]{high voltage direct current}
	\acro{ode}[ODE]{ordinary differential equation}
	\acrodefplural{pde}[PDEs]{partial differential equations}
	\acro{pde}[PDE]{partial differential equation}	
	\acrodefplural{pec}[PECs]{perfect electric conductors}
	\acro{pec}[PEC]{perfect electric conductor}
	\acrodefplural{qoi}[QoIs]{quantities of interest}
	\acro{qoi}[QoI]{quantity of interest}
	\acro{rms}[rms]{root mean square}
	\acro{sir}[SiR]{silicone rubber}
	\acro{xlpe}[XLPE]{cross-linked polyethylene}
\end{acronym}

\maketitle
\section{Introduction}\label{sec1}
When developing electrical equipment, engineers optimize initial design proposals by carefully identifying a large number of design parameters. In doing so, they rely on rules of thumb, know-how and previous experience, existing standards and, increasingly, simulation and optimization tools.
Numerical optimization is used to simultaneously improve -- possibly conflicting -- \acp{qoi}, robustness and costs. While stochastic optimization plays a major role, derivative-based deterministic optimization algorithms are becoming, again, increasingly interesting \cite{Ion_2018aa}. Their advantages over stochastic methods are a faster convergence, i.e.~less expensive optimization runs, and efficient coupling with mesh refinement and reduced order models. However, in case of derivative-based approaches, the problem of efficient gradient computation arises. 
The most common methods for gradient computation, e.g. finite differences and the \ac{dsm}, are not well suited for applications with many design parameters because their computational costs scale with the number of parameters \cite{Li_2004aa, Nikolova_2004aa}. The \ac{am}, on the other hand, has computational costs that are almost independent of the number of parameters \cite{Li_2004aa, Cao_2003aa, Nikolova_2004aa}.
The \ac{am} has previously been applied in the analysis of electric networks. 
The first formulation for the \ac{am} in this context, which was based on Tellegen's theorem \cite{Tellegen_1952aa}, was published by Director and Rohrer \cite{Director_1969aa}. Only since the 2000s, the \ac{am} has been applied to electromagnetic problems more often and remains an active field of study \cite{Nikolova_2004aa, Georgieva_2002aa, Bakr_2014aa,Park_1996aa,LalauKeraly_2013aa, Lee_2015aa,Sayed_2018aa}.
In the field of \ac{hv} engineering, Zhang et. al. recently used the \ac{am} for topology optimization of a station class surge arrester model with linear media at steady state \cite{Zhang_2021aa}. 
However, many \ac{hv} devices are exposed to transient overvoltages and contain strongly nonlinear materials, so that an investigation of the steady state alone, i.e. in the frequency domain, is not sufficient \cite{Hussain_2017aa, Spack_2021aa}. Therefore, in this work, the \ac{am} is formulated and solved numerically for the nonlinear transient \ac{eqs} problem. Additionally, a method for sensitivity calculation of \acp{qoi} evaluated at a given point in time is presented, since the \ac{am} naturally only considers time-integrated \acp{qoi}. The \ac{am} is validated using an analytical example. Subsequently, a nonlinear resistively graded 320\,kV \ac{hvdc} cable joint under impulse operation serves as a prominent technical example. It is shown that the \ac{am} is capable of computing the sensitivities of this highly transient nonlinear problem with reasonable computational effort. This is an important step towards gradient-based optimization of electric devices in \ac{hv} engineering.

\section{Electroquasistatic Problem}
The \ac{eqs} problem in time domain reads
\begin{subequations}\label{eq:joint:eqspot}
\begin{alignat}{3}
  \label{eq:eqspde}-\divergence{\sigma \grad \phi} - \divergence{\partial_t\left(\varepsilon \grad \phi\right)}&=0
  && \quad t\in[0,T],\,\,\zr \in \Omega \qscn\\
  \phi &=\phi\idx{fixed}
  && \quad t\in[0,T],\,\,\zr \in \Gamma\idx{e} \qscn\\
  -(\sigma\grad\phi + \partial_t\left(\varepsilon \grad \phi\right)\cdot\zn &=0
  && \quad t\in[0,T],\,\,\zr \in \Gamma\idx{m} \qscn\\
   \phi &=\phi_0 
     && \quad t=0\phantom{[,T]},\,\,\zr \in \Omega \label{eq:eqs:init}\qcm
\end{alignat}
\end{subequations}
where  $t$ is the time, $\zr$ is the position vector, $\Omega$ is the computational domain and $T$ is the terminal simulation time. The electric scalar potential is $\phi$. $\sigma$ and $\varepsilon$ represent the electric conductivity and permittivity, respectively. $\phi\idx{fixed}$ are the fixed voltages at the electrodes, $\Gamma\idx{e}\neq\emptyset$, and $\zn$ is the unit vector at the magnetic boundaries, $\Gamma\idx{m}=\partial\Omega\backslash\Gamma\idx{e}$. The initial condition is denoted by $\phi_0$. 
In case of a field-dependent conductivity or permittivity, i.e.~$\sigma = \sigma( E(\bm r, t), \bm r)$ and $\varepsilon = \varepsilon(E(\bm r, t), \bm r)$, \eqref{eq:joint:eqspot} becomes nonlinear.

The standard \ac{2d} axisymmetric \ac{fe} problem of \eqref{eq:joint:eqspot} is formulated by discretizing $\phi(\bm r, t) \approx \sum_j u_j N_j$,
where $N_j(\bm r)$ are linear nodal \ac{fe} shape functions. The degrees of freedom are $u_j(t)$, which are assembled in the vector $\mathbf{u}$. The semi-discrete version of \eqref{eq:joint:eqspot} according to the Ritz-Galerkin procedure reads
\begin{align}
 \mathbf{K}_\sigma \mathbf{u} + \partial_t\left(\mathbf{K}_\varepsilon \mathbf{u}\right) = 0,
 \label{eq:fe_system}
\end{align}
with 
\begin{alignat}{2}
   [\bm K_{\sigma}]_{ij} &=\int_\Omega \sigma     \grad {N_j}\cdot\grad{N_i}\udV &&\quad\quad i,j = 1,...,N_\+{N}\qcm\\
    [\bm K_{\varepsilon}]_{ij} &=\int_\Omega \varepsilon     \grad {N_j}\cdot\grad{N_i}\udV &&\quad\quad i,j = 1,...,N_\+{N}\qcm
 \label{eq:K_2D}
\end{alignat}
where $N_\+{N}$ denotes the number of nodes. For the time discretization, the implicit Euler time stepping scheme is used. The Newton method is applied in every time step to handle the material nonlinearities.

\section{Adjoint Method for Nonlinear EQS Problems}
Numerical optimization studies the effects of multiple design parameters, $\bm p =[p_1,...,p_j,...,p_{N_\+{P}}]$, on the \acp{qoi}, $G_k(\phi, \bm p)$, $k = 1,...,N_\+{QoI}$. In each \ac{fe} simulation, one parameter combination $\bm p_0$ is adopted, and the \acp{qoi} are analyzed by post processing the electric scalar potential. Common design parameters are, in particular, material parameters and the dimensions of the geometry. Taking the cable joint of section~\ref{sec:joint} as an example, possible \acp{qoi} are, e.g., the maximum tangential field stress at material interfaces, or the electric losses during impulse operation \cite{Spack_2021ab, Hussain_2017aa}. 

The \ac{am} is a method for gradient or sensitivity calculation, which is particularly efficient when the number of parameters, $N_\+{P}$, is significantly larger than the number of QoIs, $N_\+{QoI}$ \cite{Cao_2003aa, Li_2004aa}.
Sensitivities describe how and how strong a given \ac{qoi} $G_k$ is affected by a design parameter $p_j$, i.e. 
\begin{equation}
\dd {G_k} {p_j}(\bm p_0) = \pd{G_k}{p_j}(\bm p_0) + \pd{G_k}{\phi}\dd{\phi}{p_j}(\bm p_0)\qcm
\end{equation}
where $\bm p_0$ is the active parameter configuration.
In case of nonlinear media, the sensitivity of the electric potential with respect to the parameter, $\dd{\phi}{p_j}$, is typically unknown. The idea of the \ac{am} is to avoid the computation of $\dd{\phi}{p_j}$ by a clever modification of the \acp{qoi} \cite{Li_2004aa, Cao_2003aa}: The \acp{qoi} are expressed in terms of a functional $g_k$, which is integrated over the temporal and spatial computational domain, $\left[ 0,T \right] \times \Omega$. Additionally, the nonlinear \ac{eqs} problem \eqref{eq:joint:eqspot} is embedded, multiplied by a test function $w_k(\bm r, t)$, i.e. 
\begin{align}
\label{eq:qoi:fctl}
G_k(\phi,\bm p) &= \int_0^T\int_\Omega g_k(\phi,\zr,t,\bm p)\udV\,\+dt \\ 
\nonumber&\phantom{=} -  \int_0^T\int_\Omega w_k(\bm r, t)\cdot \underbrace{\left(-\divergence{\sigma \grad \phi} - \divergence{\partial_t\left(\varepsilon \grad \phi\right)}\right)}_{= 0}\udV\,\+dt\qpt
\end{align}
For any $\phi$ solving \eqref{eq:joint:eqspot}, the additional term is zero and the test function can be chosen freely. The goal of the \ac{am} is to choose the test function in such a way, that the sensitivity of the extended \ac{qoi} After a lengthy derivation, it can be shown that the unknown term is eliminated if the test function is chosen as the so-called adjoint variable, i.e. the solution of the adjoint problem \cite{Li_2004aa, Cao_2003aa}. The adjoint problem for \ac{eqs} problems with nonlinear materials reads

   \begingroup \small
  \begin{subequations}
   \label{eq:eqs:adj}
   \begin{alignat}{2}
     \label{eq:eqsadjmain} -\divergence{{\bm \sigma_d} \grad {w_k}} + \divergence{{\bm \varepsilon_d}\, \partial_t\left(\grad{ {w_k} }\right)}&=\dd {g_k} \phi\qcm\quad&& t \in [0,T],\,\, \zr \in\Omega\qscn\\
           \label{eq:eqsadjdir}{w_k} &=0\qcm&&t \in [0,T],\,\, \zr \in\Gamma_\+{e}\qscn\\
       \label{eq:eqsadjneu} -(\bm \sigma_d \grad w_k - \bm \varepsilon_d \partial_t\left( \grad w_k\right)\cdot\zn &=0\qcm&&t \in [0,T],\,\, \zr \in\Gamma_\+{m}\qscn\\
      \label{eq:eqsadjinit}{w_k}&=0\qcm && t  = T \phantom{[,0]}, \,\,\zr \in\Omega \qcm
     \end{alignat}
   \end{subequations}
\endgroup
where all quantities are evaluated at the active parameter configuration $\bm p_0$. 
Note the plus sign in front of the term with the time derivative in \eqref{eq:eqsadjmain} instead of the minus sign in   \eqref{eq:eqspde} and the terminal condition \eqref{eq:eqsadjinit} instead of the initial condition \eqref{eq:eqs:init}, which indicate that the adjoint problem needs to be integrated backwards in time or the time reversing variable transformation $\tilde t = T - t$ must be applied \cite{Cao_2002aa}.

The adjoint problem is a linear \ac{pde} that naturally includes the tensorial material linearizations $ \bm \sigma_\+d =  \dd{\bm J}{\bm E}$ and $\bm \varepsilon_\+d =\dd{\bm D}{\bm E}$. Through that, it implicitly depends on the solution of the \ac{eqs} problem, i.e. $ \bm \sigma_\+d(E)$ and $ \bm \varepsilon_\+d(E)$. Therefore, in order so solve the adjoint problem in backward mode, the \ac{eqs} problem must first be solved conventionally, i.e. in forward mode, and its solution stored for all time steps. In case of \ac{fe} simulations, this can lead to a significant memory overhead \cite{Cao_2002aa,Cyr_2014aa}. For strategies on how to reduce the memory requirement, see for example \cite{Cao_2002aa,Cyr_2014aa}.

 Once the solution of the electric potential and all adjoint variables, $w_k$, are available, all sensitivities can be computed directly by
 \begin{equation}
\begin{split}
\label{eq:eqs:adj_sens_final}
\dd{G_k}{p_j}(\bm p_0) &= {\int_0^T\int_\Omega\pd{g}{p_j} + \grad {w_k} \cdot \pd \zJ {p_j} - \grad{\pd {w_k} t} \cdot \pd{\bm D} {p_j} \udV \+dt}\\
&\phantom{=}- \intV{\grad {w_k} \cdot { \dd{\bm D}{p_j}}}\bigg\vert_{t = 0}\qcm
\end{split}
\end{equation}
where the derivative ${\dd{\bm D}{p_j}}({t = 0}) $ is obtained by differentiating the initial condition \eqref{eq:eqs:init}.
Again, all quantities are evaluated for the active parameter configuration $\bm {p}_0$. 

\subsection{Finite Element Discretization}
\label{ch:jstat:fem}
The derivative of the electric scalar potential to the parameter $p_j$ and the adjoint variable are discretized using linear \ac{fe} nodal shape functions, i.e. 
\begin{align*}
  \dd{\phi}{p_j}(\bm r, t)  &\approx \sum_{r = 1}^{N_\+{node}} u'_r(t)  N_r(\zr) &  w(\bm r, t)\approx \sum_{r = 1}^{N_\+{node}} w_r(t) N_r(\zr)\qcm
\end{align*}
and the time axis is discretized using $N_t$ samples, i.e. $t \in \{t_1 = 0,...,t_n,...,t_{N_t} = T\}$. The semi-discrete version of the adjoint problem \eqref{eq:eqs:adj} then reads
\begin{align}
\label{eq:eqs:adj:discr}
  \bm K_{\bm \sigma_d} \bm w &- \bm K_{\bm \varepsilon_d}\dd{\bm w}{t}  =\bm q \qcm
\end{align}
with
\begin{align}
  \label{eq:Kepsd}
   [\bm K_{\bm \sigma_d}]_{rs} &=\int_\Omega \grad{ N_r}\cdot\bm\sigma_d\cdot\grad{ N_s}\udV& r,s&= 1,...,N_\+{N} \qscn\\
   [\bm K_{\bm \varepsilon_d}]_{rs} &=\int_\Omega \grad{ N_r}\cdot\bm\varepsilon_d\cdot\grad{ N_s}\udV& r,s&= 1,...,N_\+{N} \qscn\\
  \label{eq:def:q} [\bm q]_r &=  \intV{\pd{g}{u_r}} & r& = 1,...,N_\+{N} \qpt
\end{align}
Finally, the semi-discrete version for the sensitivity calculation reads 
\begin{equation}
\begin{split}
\label{eq:eqs:adj:sens:disc}
  \dd{G_k}{p_j}(\bm p_0) =& \intT{\int_\Omega  \pd{g}{p_j}\udV -\bm u^T \bm K_{\sigma_p}w +\bm u^T \bm K_{\varepsilon_p} \pd w t } \\
  &\phantom=+ \bm u^T \bm K_{\varepsilon_p} \bm w \bigg\rvert_{t=0} + (\bm u')^T \bm K_{\bm \varepsilon_d} \bm w \bigg\rvert_{t=0} \qcm
\end{split}
\end{equation}
with 
\begin{align}
  \label{eq:Ksigmap}
   [\bm K_{ \sigma_p}]_{rs} &=\int_\Omega \pd{\sigma}{p_j}\grad{ N_r}\cdot\grad{ N_s}\udV& r,s&= 1,...,N_\+{N} \qscn\\
   [\bm K_{\varepsilon_p}]_{rs} &=\int_\Omega \pd{\varepsilon}{p_j}\grad{ N_r}\cdot\grad{ N_s}\udV& r,s&= 1,...,N_\+{N} \qscn\\
  \label{eq:def:q} [\bm q]_r &=  \intV{\pd{g}{u_r}} & r& = 1,...,N_\+{N} \qpt
\end{align}
In the scope of this work, the time integral of \eqref{eq:eqs:adj:sens:disc} is computed using trapezoidal integration and the time derivative in \eqref{eq:eqs:adj:discr} is approximated using the implicit Euler method.

\subsection{Treatment of pointwise QoIs}\label{sec:ex}
As can be seen from \eqref{eq:qoi:fctl}, the \ac{am} is naturally suited for integrated \acp{qoi}. Often, however, we wish to analyze \acp{qoi} that are evaluated at certain points in space or time. The evaluation at a certain position or time can be expressed by Dirac delta functions inside the functional $g_k$. To illustrate the effects this has on the \ac{am}, the electric potential evaluated at a specified position $\zr_\+{ref}$ and time $t_\+{ref}$ is considered as an example, i.e.\ 
\begin{equation}
  \label{eq:ex:phiref}   G_k= \int_0^T\int_\Omega g_k\udV\+dt = \int_0^T\int_\Omega \delta(\zr - \zr_\+{ref})  \delta(t - t_\+{ref}) \phi\udV\+dt \qcm
\end{equation}
where $\delta$ is the Dirac delta function. The \acl{rhs} of the adjoint problem is then given by
\begin{align}
 \dd {g} \phi &= \delta(\zr-\zr_0) \delta(t - t_\+{QoI})\qcm
\end{align}
and after discretization in space one finds
  \begin{align}
 \label{eq:ex:ds} \bm q(t) &=\left[\begin{array}{ccccccc} 0 & \ldots & 0 & 1 & 0 & \ldots & 0 \end{array}\right]^T  \delta(t - t_\+{QoI}) \qpt
\end{align}
In \eqref{eq:ex:ds}, the spatial integration during the derivation of the \ac{fe} formulation has converted $\delta(\zr-\zr_0) $ into a unit excitation at the corresponding node. The temporal Dirac function $\delta(t-t_\+{QoI})$ on the other hand must be approximated during numeric integration. In the context of this work, this is done by hat functions with an area of one, i.e.,
  \begin{equation}
\label{eq:diracapprox}\bm q(t_n) =\left[\begin{array}{ccccccc} 0 & \ldots & 0 & 1 & 0 & \ldots & 0 \end{array}\right]^T \,\frac 1 {\Delta_\+{imp}}\delta_{n_\+{ref}}^n\qcm
  \end{equation}
 where $\Delta_\+{imp}$ denotes the time step size right before and after $t_\+{ref}$. The approximation of the Dirac impulse with the help of other functions, e.g. a normal distribution, led to similar results. However, the approximation by a hat function is easy to implement and has the clear advantage of a compact support.
 
\section{Results}
In this section, the \ac{am} \eqref{eq:eqs:adj} for transient \ac{eqs} problems is validated. In the first step, the layered resistor of Fig.~\ref{fig:jstat:box} is considered, and the \ac{fe} adjoint and analytic sensitivities are compared. In a second step, the method is applied to a nonlinear 320\,kV cable joint specimen and the results are validated using results obtained by the \ac{dsm} as a reference.

\subsection{Analytical Example}
\begin{figure}[tbh]	
	\centering
	\hspace*{-4mm}
	\subfloat[]{\setlength{\fwidth}{0.1 \linewidth}
	\label{fig:jstat:box}\includegraphics{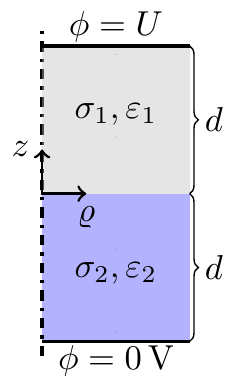}}
	\subfloat[]{\setlength{\fwidth}{0.1 \linewidth}
	\label{fig:box:phiref}\includegraphics[width=0.4 \linewidth]{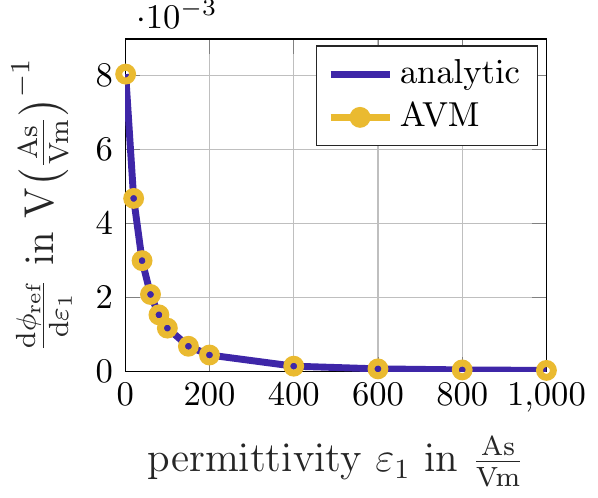}}
	\subfloat[]{\setlength{\fwidth}{0.1 \linewidth}
	\label{fig:box:relErr:phiref}\includegraphics[width=0.4 \linewidth]{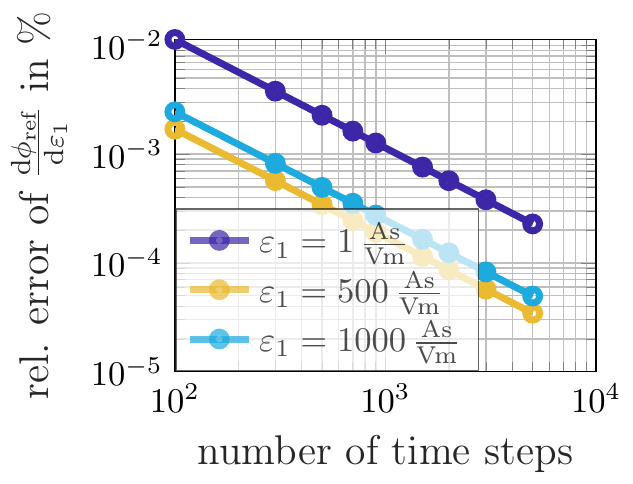}}
	\caption{(a) Resistor with two material layers of thickness $d = 1\,$cm. (b) Sensitivity $\dd {\phi_\+{ref}}{\varepsilon_1}$ for different values of $\varepsilon_1$. (c) Relative error of $\dd {\phi_\+{ref}}{\varepsilon_1}$ in \% for different numbers of time steps. }
	\label{fig:jstat:Ploss}
\end{figure}
The first example is the layered resistor depicted in Fig.~\ref{fig:jstat:box}. The upper electrode  is excited with a sinusoidal voltage, i.e., $U(t) = 1\,\+V\cdot\+{sin}(\omega t)$ with $\omega = 2 \pi 50\,\+{Hz}$, and the bottom electrode is grounded. For $t = 0$ the potential is assumed to be zero everywhere. The conductivities, $\sigma_1 = 10\,$A/Vm and $\sigma_2 = 20$\,A/Vm, and the permittivities, $\varepsilon_1 = 40$\,As/Vm and $\varepsilon_2 = 60$\,As/Vm, of the two materials are constant.
The \ac{eqs} \ac{am} is validated for an integrated \ac{qoi} as well as for non-integrated \ac{qoi}. More specifically, the \acp{qoi} are the electrical energy converted in the time span $[0\,\+s,\frac{2\pi}{\omega}\,\+s]$, 
\begin{align}
\label{eq:eqs:eloss}
  W_\+{el} &= \int_{0\,\+s}^{\frac{2\pi}{\omega}\,\+s}\int_\Omega \sigma(\Delta \phi)^2\udV\,\+dt \qcm
\end{align}
and the potential in the middle of the upper material, i.e., $\zr_\+{QoI} = (0,\frac d 2)$, evaluated at $t_\+{QoI}  = \frac{\pi}{2\omega}\,\+s$,  
\begin{align}
\label{eq:eqs:phiref}
  \phi_{\+{ref}}  &= \int_{0\,\+s}^{2\pi\,\+s}\int_\Omega \delta(\zr - \zr_\+{QoI})  \delta(t - t_\+{QoI}) \phi \udV\,\+dt \qpt
\end{align} 

First, the sensitivity $\dd {\phi_\+{ref}}{\varepsilon_1}$ of the reference potential with respect to the permittivity of the upper material is computed. The time axis is discretized using the step size $\Delta_\+{main}$. Directly before and after $t = t_\+{QoI}$ the step size is reduced to $\Delta_\+{imp} = 10^{-8}\Delta_\+{main}$ in order to approximate the Dirac impulse. Fig.~\ref{fig:box:phiref} shows that the \ac{am} is able to reproduce the analytic results of $\dd {\phi_\+{ref}}{\varepsilon_1}$ for a wide range of permittivity values. In Fig.~\ref{fig:box:relErr:phiref} a first order convergence of the relative error with respect to the number of time steps can be observed, which matches the order of the implicit Euler method.

\begin{figure}[tbh]	
	\centering
	\subfloat[]{\setlength{\fwidth}{0.1 \linewidth}
	\label{fig:box:Eloss}\includegraphics[width=0.43 \linewidth]{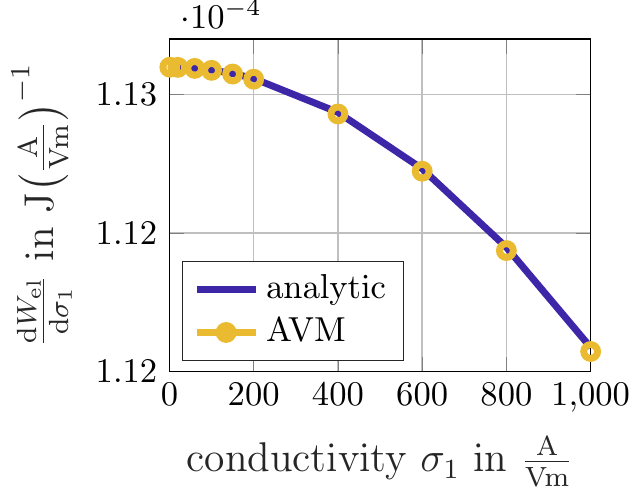}}
	\subfloat[]{\setlength{\fwidth}{0.1 \linewidth}
	\label{fig:box:relErr:Eloss}\includegraphics[width=0.43 \linewidth]{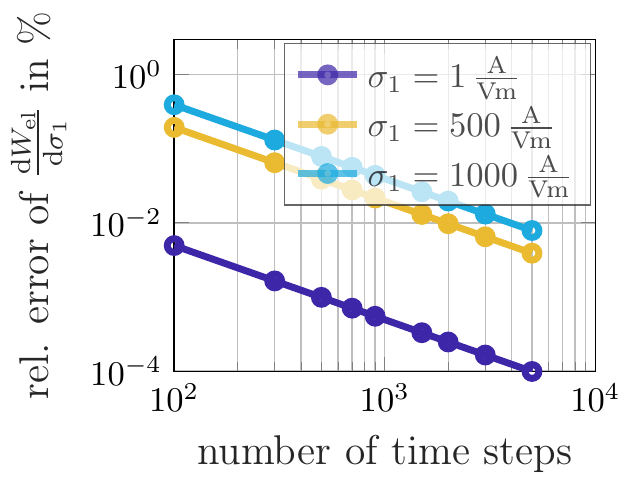}}
	\caption{(b) Sensitivity $\dd {\+ W_\+{el}}{\sigma_1}$ for different values of $\sigma_1$. (c) Relative error of $\dd {\+ W_\+{el}}{\sigma_1}$ in \% for different numbers of time steps.  }
	\label{fig:jstat:Ploss}
\end{figure}

 Next, the sensitivity $\dd {\+ W_\+{el}}{\sigma_1}$ of the electric energy with respect to the conductivity of the upper material is computed for conductivities ranging from 1\,A/Vm to 1000\,A/Vm (see~Fig.~\ref{fig:box:Eloss}). As shown in Fig.~\ref{fig:box:relErr:Eloss} the results again converge linearly with respect to the number of time steps. The \ac{am} has, thus, been successfully validated for a transient EQS problem. 

\subsection{320\,kV HVDC Cable Joint Specimen}\label{sec:joint}
The development of \ac{hvdc} cable systems is one of the greatest challenges of our time for the \ac{hv} engineering community \cite{CigreD156_2020aa, Ghorbani_2014aa, Messerer_2022aa}. 
Cable joints are known to be the most vulnerable part of \ac{hvdc} systems, as they must safely handle field strengths in the range of several kV/mm \cite{CigreD156_2020aa,Chen_2015ac,Joergens_2020ab,Poehler_2022aa}. 
The electric field stress can be reduced by adding a layer of so called \ac{fgm}, that features  a strongly nonlinear electric conductivity and balances the electric field, similar to the overvoltage clipping of metal-oxide surge arresters \cite{Spack_2016aa, Spack_2021aa}.

\begin{figure}[h]%
\centering
\includegraphics[width=0.9\textwidth]{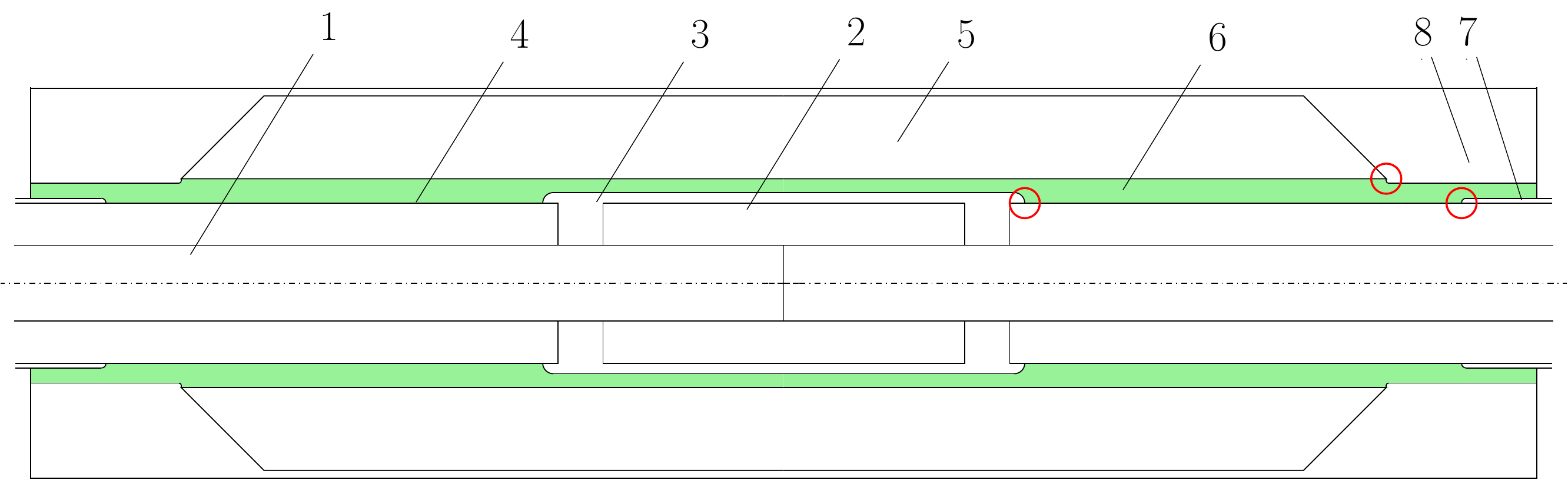}
\vspace{0.2cm}
\caption{\begingroup \small Schematic of the investigated HVDC joint in the $\varrho$-$z$-plane (drawing is not to scale). The typical positions of the maximum tangential field stresses are indicated by red circles \cite{Hussain_2017aa}. The numbers indicate the different materials as described in the text. \endgroup}\label{fig:geometry}
\end{figure}

The \ac{am} is applied to a 320\,kV \ac{hvdc} cable joint specimen, which is adopted from  \cite{Hussain_2017aa} and shown in Fig.~\ref{fig:geometry}. 
The joint connects two copper conductors (domain 1) with an aluminum connector (domain 2). These domains are covered by a layer of conductive \ac{sir} (domain 3). The cable insulation consists of \ac{xlpe} (domain 4) and the joint insulation of an insulating \ac{sir} (domain 5). Both insulation layers are separated by a nonlinear resistive \ac{fgm} (domain 6, highlighted in green). The outer conductive \ac{sir} sheaths of the cable (domain~7) and the joint (domain~8) are on ground potential. Inside the \ac{fgm}, elevated field stresses occur at the triple points, i.e., the contact points of \ac{fgm}, insulating material and conductive \ac{sir} (indicated by red circles). 
The joint is subjected to an impulse overvoltage with an amplitude of $\hat U= 100\,$kV that is superimposed on the \ac{hvdc} excitation of 320\,kV. The transient standard 1.2/50 lightning impulse is given by \cite{Kuchler_2017aa}
\begin{equation}
	\label{eq:impulse}
	U_\+{imp}(t) = \hat U \, \frac{\tau_2}{\tau_2-\tau_1}\,\left( \+{exp}\left( -\frac{t}{\tau_2}\right)-\+{exp}\left( -\frac{t}{\tau_1}\right)\right)\qcm
\end{equation}
with $\tau_1 = \frac{1.2\,\upmu\+s}{2.96}$ and $\tau_2 = \frac{50\,\upmu\+s}{0.73}$. 
The excitation is applied to the conductive \ac{sir} covering the conductor and the conductor clamp, which is modeled as a perfect electric conductor.

The field-dependence of the conductivity of the \ac{fgm} is described by the analytic function
\begin{equation}
\sigma(E) = a_1 \frac{1 + a_4^{(E-a_2)\,a_2^{-1}}}{1 + a_4^{(E-a_3)\,a_2^{-1}}}\qcm
\label{eq:ftsig}
\end{equation}
with the parameters $a_1 = 10^{-10}\,\text{A/Vm}$, $a_2 = 0.7\cdot 10^{6}\,$V/m, $a_3 = 2.4\cdot 10^{6}$\,V/m and $a_4 = 1864$. 
The simulation is performed with a mesh consisting of 51681 nodes and 106113 elements.

\begin{figure}[tbh!]	
	\centering
	\subfloat[]{\setlength{\fwidth}{0.4 \linewidth}
		\label{fig:jstat:ball:sigma_vary_E1}\includegraphics[width = 0.4\textwidth]{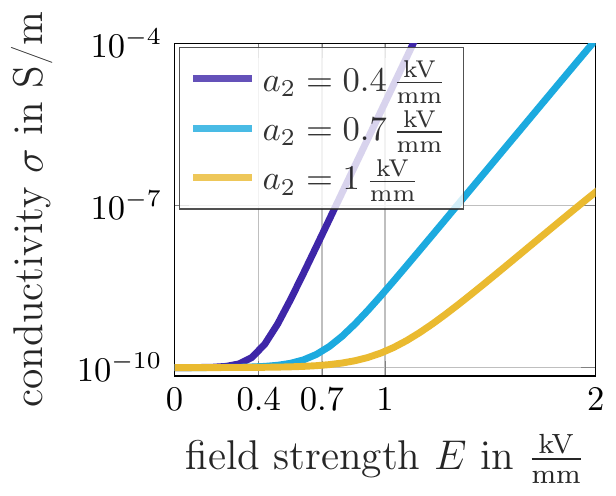}}	
		\subfloat[]{\setlength{\fwidth}{0.4 \linewidth}
		\label{fig:eqs:joint}\includegraphics[width = 0.4\textwidth]{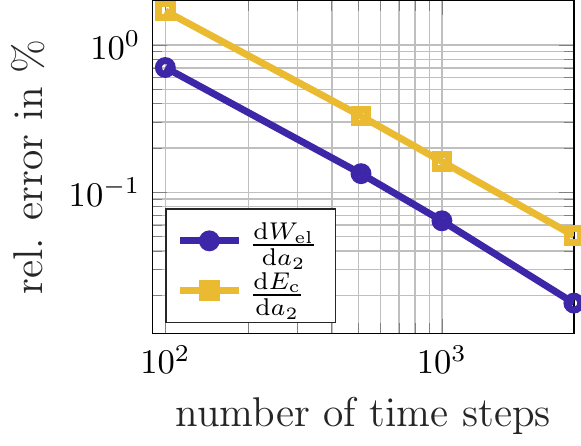}}
		\caption{(a) Field-dependence of the nonlinear conductivity defined for different values of the switching field strength, $a_2$. (b) Relative error of $\dd {W\+{el}}{a_2}$ and $\dd {E\+{c}}{a_2}$ for different numbers of time steps in \%.}
	\label{fig:eqs:validPhiref}
\end{figure}

The results of the EQS \ac{am} are validated against results of the DSM for two exemplary sensitivities. Again, both a time-integrated QoI as well as a \ac{qoi} evaluated at a specific point in time are considered, i.e., the electric losses, $W_\+{el}$, during the time span $[0\,\upmu\+s,t_\+{rise}]$ 
and the critical electric field stress, $E_\+{c}$, in the proximity of the triple point next to the conductor clamp during peak excitation. The derivatives of the \acp{qoi} are computed with respect to the switching field strength, $a_2$, which  determines when the conductivity changes from the base conductivity $a_1 = 10^{-10}$\,S/m into the strongly nonlinear region of \eqref{eq:ftsig} (see Fig.~\ref{fig:jstat:ball:sigma_vary_E1}).

Fig.~\ref{fig:eqs:joint} shows the relative error of the derivatives for different numbers of time steps. A first order convergence due to the Euler time-stepping scheme is observed for both \acp{qoi}. 
More importantly, the computational cost is reasonable for both the time-integrated \ac{qoi} as well as the \ac{qoi} evaluated at a given point in time. With only 200 time steps, the relative errors are below one percent, requiring a simulation time in the range of tens of minutes. The \ac{am} was, thus, successfully validated for a strongly nonlinear and highly transient example. It was shown that with the method presented in section~\ref{sec:ex} the \ac{am} is no longer restricted to integrated \acp{qoi}. Moreover, it was demonstrated that even for this very challenging example the computation time lies within reasonable limits, which is an important first step towards gradient-based optimization.

\section{Conclusion}
The adjoint variable method is a method for calculating gradients of selected quantities of interest with respect to a set of design parameters. It has computational costs nearly independent of the number of design parameters and is, thus, very efficient for problems where the number of parameters is larger than the number of quantities of interest.
In this work, the \acl{am} is adopted for transient \acl{eqs} problems with nonlinear material characteristics. The adjoint \acl{pde} is presented and formulated as a \acl{2d} axisymmetric finite element problem. It is shown, how to consider quantities of interest evaluated at specific points in space or time. After validating the method against an analytic example, the method is applied to a 320\,kV \acl{hvdc} cable joint featuring  a layer of nonlinear \acl{fgm} which is exposed to an impulse overvoltage. The results of the \acl{am} are validated using  the \acl{dsm} and it is shown that the computational costs of the \acl{am} are, even for this strongly nonlinear technical example, within reasonable limits. This is an important step towards gradient-based optimization of \acl{hv} equipment.

\bmhead{Competing Interests}

The authors declare no conflict of interest.

\bmhead{Acknowledgments}

The authors thank Rashid Hussain for providing the simulation model and material characteristics published in \cite{Hussain_2017aa}, and Myriam Koch for the helpful discussions on \ac{hvdc} cable joints. \\
This work is supported by the Graduate School CE within the Centre for Computational Engineering at the Technische Universität Darmstadt.

\bibliography{thesis,additions_ysl}
\bibliographystyle{sn-basic.bst}


\end{document}